%% file: main.tex
\documentclass[twocolumn, twoside, 10pt]{article}

\usepackage[
    letterpaper,
    footskip=8mm,
    headsep=8mm,
    left=2cm,
    right=2cm,
    top=3cm,
    bottom=3cm,
    ]{geometry}
  
\title{GazeGPT: Augmenting Human Capabilities using Gaze-contingent Contextual AI for Smart Eyewear}

\makeatletter
\edef\mytitle{\@title}
\makeatother

\usepackage[en-US]{datetime2}
\DTMlangsetup{showdayofmonth=false}
\usepackage{fancyhdr}
\fancypagestyle{plain}{%
  \fancyhf{}%
  \fancyfoot[R]{\footnotesize Uploaded to arXiv \today}%
}
\usepackage{booktabs} 
\usepackage[hidelinks]{hyperref}
\usepackage{graphicx}

\usepackage{libertine}
\usepackage[T1]{fontenc}
\usepackage[utf8]{inputenc}

\usepackage[
    backend=biber,
    style=authoryear,
    minnames=1,
    maxnames=2,
    uniquelist=minyear,
]{biblatex}
\makeatletter

\usepackage{url}
\newrobustcmd*{\parentexttrack}[1]{%
  \begingroup
  \blx@blxinit
  \blx@setsfcodes
  \blx@bibopenparen#1\blx@bibcloseparen
  \endgroup}

\AtEveryCite{%
  }

\makeatother
\let\cite\parencite
\newcommand{\shortcite}[1]{\parencite*{#1}}

\usepackage{enumitem}

\usepackage{caption}
\usepackage[ruled]{algorithm2e} 

\SetAlFnt{\small}
\SetAlCapFnt{\small}
\SetAlCapNameFnt{\small}
\SetAlCapHSkip{0pt}

\usepackage{authblk}
\makeatletter
\renewcommand\AB@affilsepx{, \protect\Affilfont}
\makeatother

\renewcommand\Affilfont{\sf}
  
\usepackage{sectsty}
\sectionfont{\large\sf\MakeUppercase}
\subsectionfont{\sf}

\usepackage{titlesec}
\titlespacing*{\section}
{0pt}{2.5ex plus 0.5ex minus .2ex}{0.5ex plus .2ex}
\titlespacing*{\subsection}
{0pt}{2.5ex plus 0.5ex minus .2ex}{0.5ex plus .2ex}

\makeatletter
\def\@maketitle{%
  \newpage
  \begin{flushleft}%
  \let \footnote \thanks
  \sf
    {\LARGE \@title\par}%
   \vskip 1.5em%
   {\large
     \lineskip .5em%
       \@author
       \par}%
   \vskip 1em%
   {\large \@date}%
  \end{flushleft}%
  \par
\begin{center}
  \includegraphics[width=\textwidth]{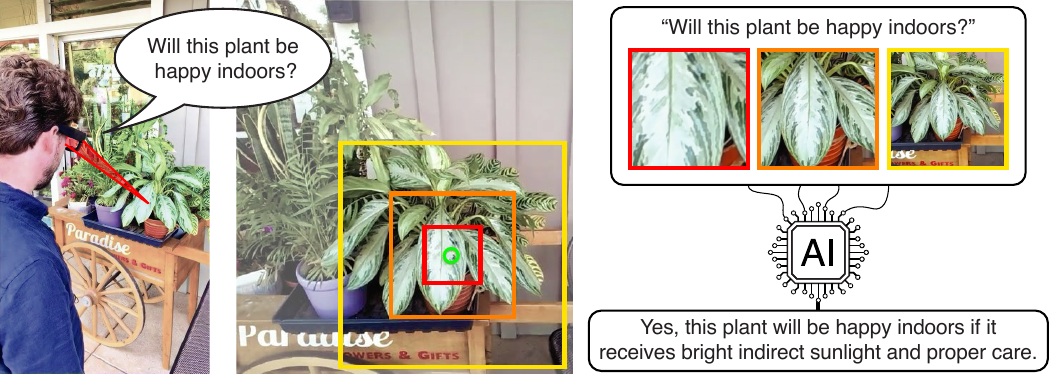}
  \captionof{figure}{We introduce GazeGPT, a human-centric interface to generative AI models. Current AI models are exceptional at ingesting multimodal data and providing reasonable responses, but often lack the fundamental information to identify the object of interest to the human user. GazeGPT uses a combination of a gaze tracker and a world-facing camera to provide context to user queries. The query, along with a multiscale crop around the object of interest, is uploaded to a multimodal large language model, like GPT-4V, which can provide better responses with the included context. This new interface to AI has the potential to fundamentally change how humans access information.}
  \label{fig:teaser}
\end{center}
}
\makeatother

\DeclareRobustCommand{\thinskip}{\hskip 0.1em\relax}
\def\emdash{---}
\def\d@sh#1#2{\unskip#1\thinskip#2\thinskip\ignorespaces}
\def\Dash{\d@sh\nobreak\emdash}

\newcommand{\auth}[2][]{\author[#1]{\MakeUppercase{#2}}}

\begin{document}

\auth[1]{Robert Konrad}
\auth[1]{Nitish Padmanaban}
\auth[1]{J. Gabriel Buckmaster}
\auth[1]{Kevin C. Boyle}
\auth[1,2]{Gordon Wetzstein}
\affil[1]{Zinn Labs, Inc.}
\affil[2]{Stanford University}
\date{}

\maketitle

\begin{abstract}
Multimodal large language models (LMMs) excel in world knowledge and problem-solving abilities. Through the use of a world-facing camera and contextual AI, emerging smart accessories aim to provide a seamless interface between humans and LMMs. Yet, these wearable computing systems lack an understanding of the user's attention. We introduce GazeGPT as a new user interaction paradigm for contextual AI. GazeGPT uses gaze tracking to help the LMM understand which object in the world-facing camera view a user is paying attention to. Using extensive user evaluations, we show that this gaze-contingent mechanism is a faster and more accurate pointing mechanism than alternatives, that it augments human capabilities by significantly improving their accuracy in a dog-breed classification task, and that it is consistently ranked as more natural than head- or body-driven selection mechanisms for contextual AI. Moreover, we prototype a variety of application scenarios that suggest GazeGPT could be of significant value to users as part of future AI-driven personal assistants.
\end{abstract}

\section{Introduction}
\label{sec:intro}
\input{sections/1_introduction}

\section{Related Work}
\label{sec:related}
\input{sections/2_related}

\section{GazeGPT System}
\label{sec:system}
\input{sections/3_system}

\section{Experiments}
\label{sec:experiments}
\input{sections/4_experiments}

\section{Applications}
\label{sec:applications}
\input{sections/5_applications}

\section{Discussion}
\label{sec:discussion}
\input{sections/6_discussion}

\AtNextBibliography{\small}
\printbibliography

\begin{figure*}[p!]
    \centering
    \includegraphics[width=\textwidth]{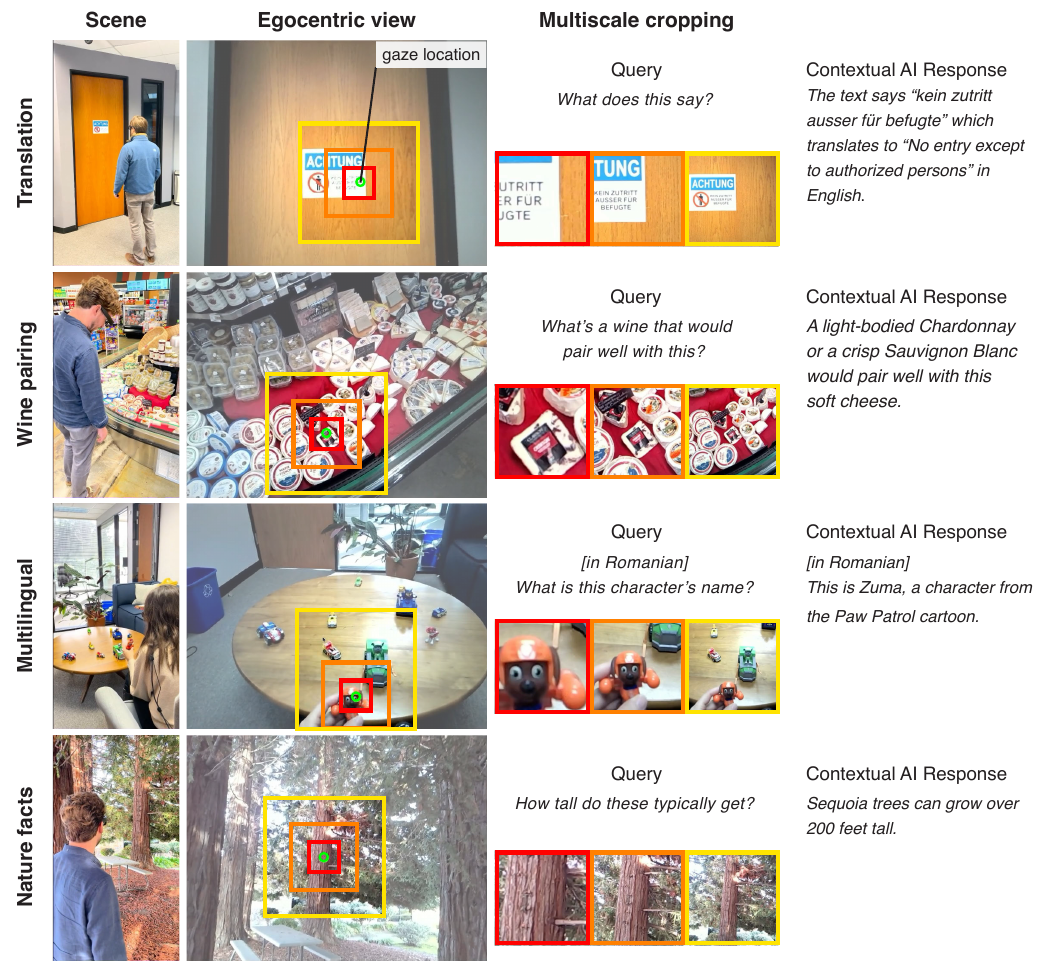}
    \caption{Applications of GazeGPT. The GazeGPT system excels at a wide variety of tasks, including general knowledge, contextual recommendations, translation, and even operates in multiple languages.}
    \label{fig:gazegpt_applications}
\end{figure*}

\end{document}

%% file: sections/1_introduction.tex
The recent emergence of multimodal generative artificial intelligence (AI)
is heralded as a significant step towards artificial general intelligence~\cite{openai2023gpt4,team2023gemini,moon2023anymal,you2023ferret,liu2023llava,liu2023improvedllava}. Indeed, some of these models already outperform human experts on world knowledge and problem-solving abilities across multiple subjects, such as math, physics, history, law, medicine, and ethics~\cite{team2023gemini}. These large multimodal models (LMMs) represent the foundation of future personal assistants that use contextual AI to understand a user's intent and augment human capabilities.

One of the key questions in this context is what an effective interface between an LMM and a human should look like. Today's models primarily take a user-defined text prompt and a specified image as input and generate text output. 
Through the use of world-facing cameras, microphones, open-ear headphones, and speech-to-text capabilities, emerging ``smart'' glasses and other accessories provide a more seamless, all-day wearable interface between humans and LMMs. Examples include Ray-Ban Meta Stories, Amazon Echo Frames, Snap Spectacles, and Humane's AI Pin, among others. Unlike conventional augmented- or mixed-reality systems, however, smart accessories usually do not use a display for visual feedback. 

The core problem with smart accessories is that LMMs have little to no understanding of a user's visual spatial attention, hampering their understanding of user intent. Imagine a cluttered environment in which a user asks the question, ``What is this?'' The best guess of which object the user is referring to may be the one centered in the camera image; however, the optical axis of a body- or head-mounted camera does not necessarily coincide with the user's attention.

Gaze is an intuitive selection mechanism and has been shown to be more effective than other ways of pointing in certain conditions (see Sec.~\ref{sec:related}). Motivated by this insight, we introduce GazeGPT as a new user interaction paradigm for contextual AI. Contextual AI uses multimodal input, including text, images, and video, to understand a user's intent and enhance their capabilities. GazeGPT adds another modality to the AI's input: the visual spatial attention of the user as indicated by their point of gaze. The point of gaze is measured by an eye tracker and used in conjunction with world-facing cameras to identify the object or region that the user fixates. A multiscale image crop centered around this point is provided to the LMM as input. To evaluate the idea of gaze-contingent contextual AI and assess its effectiveness, we build a prototype wearable-computing evaluation platform that includes a high-resolution, wide-field-of-view, world-facing camera, a microphone and speech-to-text capabilities, a GPT-4V(ision) LMM backend, and text-to-speech output.
Moreover, we design and conduct a number of user studies. These studies demonstrate that, in the absence of visual feedback, gaze is a faster and more accurate selection mechanism than body- or head-based alignment. This insight indicates that GazeGPT may be a more effective interface for contextual AI than current solutions. We also show that our GazeGPT system can close the gap in task performance between humans and generative AI. Specifically, GazeGPT enables a human to achieve close-to-AI-level performance in difficult object classification tasks. Finally, users consistently rank a gaze-directed interface to an LLM as more natural than head- or body-directed interfaces in our study.

GazeGPT is an example of human-centric AI: a technology that enhances, not replaces, people's natural abilities. We show that it provides an intuitive interface between humans and LMMs, representing a crucial part of future contextual AI and smart eyewear.

Specific contributions of our work include:
\begin{itemize}[noitemsep]
    \item the introduction of the gaze-contingent contextual AI paradigm along with a prototype platform;
    \item an evaluation of selection mechanisms, showing that gaze is faster and more accurate than body- and head-based selection without visual feedback;
    \item the demonstration of augmented human capabilities, improving object classification accuracy of humans to a level that closely matches the performance of an LMM.
\end{itemize}

%% file: sections/2_related.tex
Our work is at the convergence of two seemingly disparate fields: large multimodal models and gaze-contingent graphics techniques. Here, we briefly review the most closely related work in both areas. 

\paragraph{\textbf{Multimodal Large Language Models (LMMs)}} A comprehensive overview of LMMs and multimodal foundation models in general can be found in the recent survey by Li et al.~\shortcite{li2023multimodal}. These models extend the capabilities of large language models by accepting not only text but also image data as input. The performance of LMMs is typically benchmarked in a variety of tasks, including simulated exams and mathematical reasoning as well as natural image, audio, and video understanding.

\paragraph{\textbf{Gaze-based User Interfaces (UI)}} With the emergence of eye tracking technology, gaze-based human--computer interaction techniques have gained much popularity~\cite{jacob1990you,majaranta2014eye,10.1145/3491207}. For example, gaze-based selection was shown to be faster than mouse-based selection~\cite{10.1145/332040.332445} and to 
outperform head-based selection in terms of speed, task load, required head movement, and user preference~\cite{10.1145/3206343.3206349} when used as an interface with a monitor. In the absence of visual feedback, gaze-based selection was also shown to be faster than hand-based pointing~\cite{10.1145/332040.332443}, although this trend seems to be reversed with visual feedback~\cite{cournia2003gaze}.  Because gaze is such an intuitive and effective selection mechanism, it is the primary means by which users select objects and UI elements in Apple's upcoming Vision Pro Mixed Reality Headset.

\paragraph{\textbf{Eye Tracking and Smart Eyewear}}
Eye tracking and gaze-con\-tin\-gent graphics techniques beyond UI applications are a core aspect of emerging augmented- and virtual-reality (AR/VR) systems. For example, eye tracking in AR/VR is routinely used for foveated rendering~\cite{friston2019perceptual,geisler1998real,guenter2012foveated,kaplanyan2019deepfovea,patney2016towards,tariq2022noise,tursun2019luminance,krajancich2023towards}, varifocal display techniques~\cite{akcsit2017near,7829412,liu2008optical,padmanaban2017optimizing}, enhancing depth perception~\cite{konrad2020gaze,krajancich2020optimizing}, vision correction~\cite{padmanaban:autofocals}, and many other applications surveyed by Duchowski et al.~\shortcite{duchowski2004gaze} and Koulieris et al.~\shortcite{koulieris2019near}. 

The work closest to ours is perhaps Meta's Project Aria~\cite{somasundaram2023project}: a display-less smart glasses prototype equipped with several microphones and sensors, including eye trackers and world-facing cameras. While the capabilities of this hardware platform are somewhat similar to our evaluation platform, the core contribution of our work is not a specific hardware implementation but the introduction and evaluation of the gaze-contingent contextual AI paradigm. To the best of our knowledge, this idea has not been discussed or evaluated in prior work.

%% file: sections/3_system.tex
GazeGPT is a hardware and software system that captures a world-facing image, the user's gaze, and the user's query and then submits these data to a LMM for processing and response.  

\subsection{User Interface}
\label{sec:user_interface}
A GazeGPT query begins with a button press. At that moment, the system captures the user's gaze and an image from the world-facing camera. The two are combined to create a gaze-centric image (described below). While keeping the button held down, the user speaks their query, which is then transcribed via speech to text upon release. The query and gaze-centric world view are uploaded to GPT-4V via the OpenAI API. GPT-4V's response is synthesized into speech via a text-to-speech algorithm and played back to the user over the speaker in the frames.

\subsection{Implementation}
\label{sec:implementation}

\begin{figure}[t]
    \includegraphics[width=0.8\columnwidth]{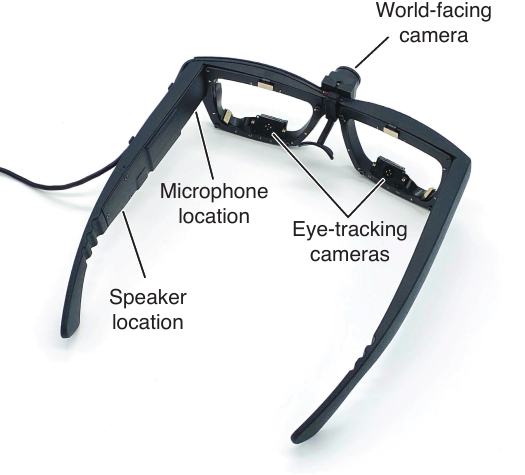}
    \vspace{-10pt}
    \caption{The Zinn Labs DK1 Evaluation Kit. The major components used in the GazeGPT system (microphone, speaker, eye tracking cameras, and world-facing camera) are labeled.}
    \label{fig:system}
\end{figure}

\paragraph{\textbf{Hardware}} 
The core hardware of the GazeGPT system is a Zinn Labs DK1 Evaluation Kit. This tethered pair of glasses includes an event-sensor-based eye tracker, world-facing camera, microphone, and speaker (Fig.~\ref{fig:system}). The eye tracker operates at 120~Hz and is rated by the manufacturer as having 1\textdegree{} accuracy and 0.4\textdegree{} precision. The glasses are tethered to a laptop computer, in our case a Razer Blade 15$''$ 2019 with an Intel I7-10750H processor, which performs the gaze estimation via the provided Zinn Labs SDK. The world-facing camera comprises a Sony IMX179 sensor integrated into a 78\textdegree{} diagonal-field-of-view module and interfaces to the tethered computer via a USB 2.0 connection. The camera captures up to 8~MP (3264 $\times$ 2448) images at 15~Hz and appears as a webcam on the laptop. The Zinn Labs DK1 provides its gaze estimate as a pupil center position and gaze vector relative to its world-facing camera. We calibrate the camera and project the 3D gaze into the 2D camera image for our analysis. A Logitech R800 Powerpoint Clicker is used as the user-interaction button.

\paragraph{\textbf{Speech transcription and synthesis}}
GazeGPT uses OpenAI's Whisper \texttt{v2-large} model for the speech-to-text transcription. For text-to-speech synthesis, the system uses ElevenLabs' Turbo V2 model. At the cost of latency, we also support ElevenLabs' Multilingual V2 model in case the transcribed speech is not English.

\paragraph{\textbf{Multimodal Large Language Model}}
By handling the text-to-speech and speech-to-text conversions separately, the GazeGPT system only requires an LMM that accepts text and visual data. The system interfaces with OpenAI's GPT-4V LMM via OpenAI's Python API by uploading text and the gaze-centric world view images encoded in base 64 format. The GPT-4V LMM returns its response as text. In the future, as LMMs become more powerful, they may be able to accept spoken audio and images, process them, and output a response as audio directly. 

\begin{figure}[t]
    \centering
    \includegraphics[width=\columnwidth]{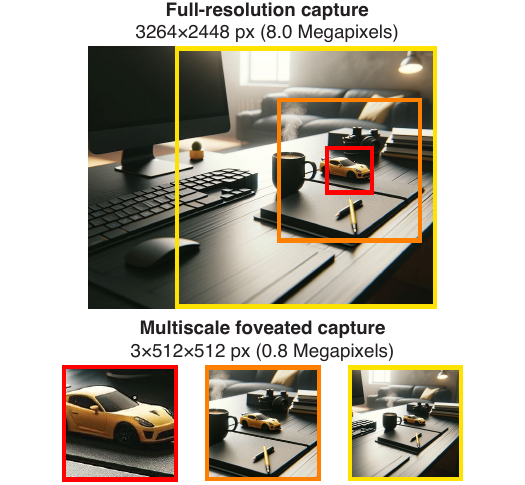}
    \caption{An illustration of the multiscale capture concept. The narrowest field of view gives a detailed view of the car that the user is looking at, while the wider field of view images provide helpful context. At the same time, the total image size is reduced by an order of magnitude.}
    \label{fig:bandwith}
\end{figure}

\paragraph{\textbf{Multiscale Cropping}} 
When the user poses a query to the Gaze\-GPT system, an 8~MP image is captured from the world-facing camera. This image is cropped and scaled to form three 512$\times$512 pixel images centered on the user’s gaze with increasingly broader fields of view, which are then uploaded to the GPT-4V model (Fig.~\ref{fig:bandwith}). These multiscale images direct GPT-4V’s attention to the object that the user is looking at and enable GPT-4V to provide useful information about objects of varying size and distance, as well as reason about objects based on their context. This technique greatly reduces data processing and transfer requirements, and results in an approximately 10$\times$ reduction in data as compared to using the full 8~MP image. 

In a system that seeks to achieve human-level visual acuity over the entire natural gaze range, the data savings may be even more dramatic. Peak human visual acuity is approximately 60~cycles per degree at the center of the fovea, which would require a digital camera system with 120~px per degree to match. Maintaining this angular resolution over the entire natural gaze range (approximately 44\textdegree{} horizontal $\times$ 33\textdegree{} vertical for the 75th percentile of gaze angles \cite{aizenman2023gaze}) plus 2\textdegree{} on each side to account for the radius of the fovea results in a 5760$\times$4320~px (24.9~MP) image. Instead, suppose a gaze-based system only captures the $\pm$2 degree region around the user's gaze at the full resolution of the image sensor. This 480$\times$480~px image can be augmented with three successively increasing fields of view that are each also scaled to 480$\times$480~px and the resulting multiscale image would be only 4$\times$480$\times$480~px (0.9 MP). This represents a savings of over 25$\times$, while still providing the LMM sufficient information to answer the user’s query.

\paragraph{\textbf{Evaluating Latency}}
System latency was measured from the end of the user question to the beginning of the spoken response. This time was broken into three parts (speech to text, GPT-4V response time, and text to speech) and measured in several typical use cases. The system was running on 200~Mbps download/20~Mbps upload business-class cable internet over WiFi. GPT-4V response time dominated overall latency at 4.0~s average, while speech to text took 0.9~s on average and text to speech took 0.3~s on average.

%% file: sections/4_experiments.tex
We evaluate gaze-tracking alongside head- and body-based selection, used by Ray-Ban Meta Glasses and Humane's AI pin respectively, to understand which, if any, of these modes can be used to augment human capabilities. To this end, we evaluate each selection mode in terms of how accurately and quickly users can select objects of interest, how accurately a system controlled by each selection mode can identify dog breeds based on images alone, and in a qualitative preference study.

Twelve adults participated (age range 18--37, 5 male) in the experiments. Participants who wore glasses for vision correction were asked to remove them for the study, but each participant was screened to make sure that they could see a 1\textdegree{} cross rendered on the screen from 1~m away. Each participant gave informed consent. 

\subsection{Evaluating Selection Mode Speed and Accuracy}
In this experiment, we aim to evaluate the emerging contextual AI selection modes \Dash gaze-, head-, and body-based \Dash in terms of their accuracy and speed of selecting objects. Specifically, we aim to compare these modes in terms of how effective they are for use in the absence of visual feedback and without training or learning effects. We compare these selection modes against smartphones, the status quo in mobile computing interfaces. While prior art exists comparing subsets of these selection modes~\cite{10.1145/332040.332445, 10.1145/3206343.3206349, 10.1145/332040.332443}, we are not aware of any work encompassing this set specifically in the absence of visual feedback \Dash a likely requirement until the emergence of all-day-wearable head-worn displays like mixed-reality headsets.

\paragraph{\textbf{Selection Mode Hardware}}
We used three different systems to evaluate the selection modes: for the gaze- and head-based modes, we used the Zinn Labs DK1; for the body-based and smartphone-like modes, we built custom devices using the same IMX179 camera as the Zinn Labs DK1 to ensure comparable image quality and resolution across modes. The body-based device consisted of a custom 3D-printed IMX179 camera mount attached to a GoPro Performance Chest Mount allowing for vertical tilt of the camera. The smartphone-like device combined a WaveShare 5.5$''$ capacitive touch AMOLED display with the IMX179 camera, mounted together via a custom 3D-printed enclosure. We streamed a live camera feed to the display, mimicking a smartphone camera app. We rendered a green circle in the center of the camera feed, spanning half the narrower display dimension. The circle serves to indicate to the user that they should center an object and additionally makes it slightly easier to do so. A picture could be taken by pressing anywhere on the display. For more details, please refer to the supplement.

\paragraph{\textbf{Study Setup and Stimuli}}
All visual stimuli were rendered on an LG UQ7590 4k 75$''$ television placed on a height-adjustable standing desk, 1~m in front of users. The height of the standing desk was adjusted such that the TV was centered at the user's eyes while they were standing. The study was performed in a dark room with overhead lighting turned off so that the main light source was the TV.

For the duration of the study, users wore the Zinn Labs DK1 system and the body-mounted GoPro chest mount. We checked that the cables running from the glasses and body-mounted system did not inhibit motion and adjusted them as necessary. Users used a Logitech R800 Powerpoint clicker to indicate intent for gaze-, head-, and body-based selection. Users were handed the phone-like device when evaluating phone-based selection. Instead of using the clicker, users tapped the touchscreen to indicate intent.

\begin{figure}[t!]
    \includegraphics[width=\columnwidth]{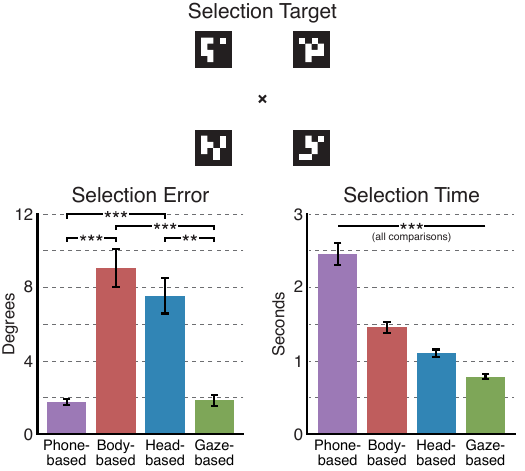}
    \caption{Results of selection evaluation for accuracy (left) and speed (right) for the selection target shown (top). Both the phone- and gaze- based selection modes achieved high accuracy (just under 2\textdegree{}), while the gaze-based selection mode was the fastest of all the modes. Significance is indicated at the **$p = 0.01$ and ***$p = 0.001$ levels. Errors bars indicate SE.}
    \label{fig:selection}
\end{figure}

The selection target consisted of a 1.06\textdegree{} cross ($\times$) that appeared black when first rendered (Fig.~\ref{fig:selection}). The cross was surrounded by 4 ArUco markers that were used to register the head/body/phone camera relative to the display to compute accuracy statistics. The cross appeared in one of 25 locations in a 5$\times$5 grid, spaced 11\textdegree{} apart and centered relative to the TV.

\paragraph{\textbf{Procedure}}
We evaluated the speed and accuracy of the selection modes with 25 trials in each mode. The order of modes and the trials were randomized, with the targets appearing once in each possible location for each mode.

At the start of each selection mode, users were instructed on how to use that mode, given the phone-like device if needed, and allowed to try a few practice selections. For the gaze-based mode, the user followed the 5-point manufacturer-provided gaze calibration immediately before the trial set to minimize eye-tracker error. They were instructed to look naturally to the object. For more details regarding the exact instructions used for each mode, please refer to the supplement. Once they understood the selection mode, they would then continue to the main set of 25 trials. 

Each trial began with a waiting screen reading, ``Press to start...'' and a central cross. They were asked to reset their selection device or gaze to the central cross \Dash or in the case of the phone, to return it to their side (to account for the fact that in a real-world scenario, their phone would likely be in their pocket or on a table). The user then tapped the interaction button or screen. The text would disappear, and after a random interval (2, 3, or 4~s), the central cross would be replaced by a test target at a randomized location. When the user selected the test target, it would first turn yellow, acknowledging their selection, and then green once the images and gaze were captured \Dash users were instructed to remain in place until the target turned green. The next trial would then begin.

Once all 25 trials for a given selection mode were completed, we proceeded to the next selection mode until all four were completed.

\paragraph{\textbf{Analysis}}
During the study, we recorded the pixel locations in the camera image of the test target (as calculated based on the ArUco markers) and the user's selection. For the gaze-based mode, the selected pixel was given by the Zinn Labs DK1; for the other modes, we took the center pixel of the camera image. These pixel locations were converted to degrees relative to the user's head based on the ArUco markers and the known location of the current target on the TV. The user's selection error was calculated as the distance between the target and selection points in degrees.

We also recorded the time taken for the user to make their selection. The timer started when the test target appeared and stopped when they selected it (i.e., when the target turned yellow).

For each user and mode, the selection error and time were averaged across all 25 trials. We analyzed the selection error averages and selection time averages each with a one-way repeated-measures analysis of variance (ANOVA) with the independent variable of selection mode having four levels. Greenhouse--Geisser sphericity correction was applied. Post hoc tests were conducted as pairwise \textit{t}-tests with Bonferroni correction applied to the \textit{p}-values \Dash reported \textit{p}-values have the Bonferroni correction factor applied.

\paragraph{\textbf{Results}}
The average selection error and time for each mode can be found in Fig.~\ref{fig:selection}. In terms of error, phone- and gaze-based selection are just below 2\textdegree{} of error, while body- and head-based are above 7\textdegree{}. On the other hand, for selection time, phone-based is the slowest at around 2.5~s, body-based is about 1.5~s, head-based $\approx$ 1.1~s, and gaze-based $\approx$ 0.8~s.

The ANOVA shows a signification effect of mode for both selection error ($F_{1.93, 21.19} = 36.12,\ p < 0.001 $) and time ($F_{1.36, 14.94} = 81.01,\ p < 0.001 $). Therefore we conduct follow-up pairwise \textit{t}-tests for the post hoc analyses.

For selection error, the \textit{t}-tests show the following significant effects: phone-based selection has lower error than body- and head-based (both $p < 0.001$), gaze-based selection has lower error than body-based ($p < 0.001$), and gazed-based selection lower error than head-based ($p < 0.01$). For selection time, all pairwise comparisons are significant ($p < 0.001$).

Overall, phone- and gaze-based selection have much lower error than the other two modes. However, gaze- and phone- based selection are differentiated by the selection time results, in which phone-based selection is slower than gaze-based selection by more than a factor of three. Body- and head-based selection have unacceptable error, at least without visual feedback. Additionally, they are also both slower than gaze-based selection, corroborating Blattgerste et al.'s~\shortcite{10.1145/3206343.3206349} finding that head-based is slower than gaze-based selection.

\subsection{Augmenting Human Capabilities}
In this experiment, we evaluate how an LMM paired with the different selection modes fares on a multimodal task. We compare the end-to-end dog-breed identification capabilities of the GazeGPT system to that of humans on the same task. We explore how the choice of selection mode affects the system's capabilities and which modes are viable for augmenting human capabilities on this task and others.

We use the same gaze-, head-, and body-based systems as in the previous experiment. We omit the phone-based mode in this experiment assuming that, with visual feedback, the human will be able to accurately select the dog of interest. We use the same study setup as in the previous study.

\paragraph{\textbf{Stimuli}}
We selected a set of 81 officially recognized dog breeds according to the American Kennel Club\footnote{\url{https://www.akc.org/}} for this task. Each dog breed had an associated 700$\times$700~px image of the dog on a white background. Of the 81 dog breeds, 25 were randomly selected to be part of the test set. 

\begin{figure}[t!]
    \includegraphics[width=\columnwidth]{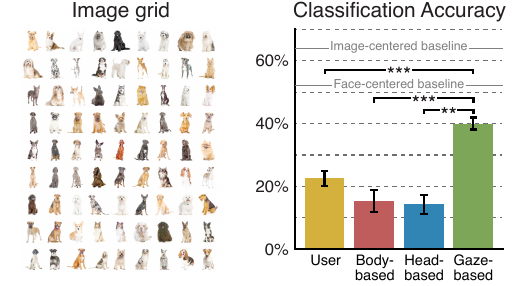}
    \caption{Example images used for the classification evaluation (left) and the results of the evaluation (right). Each trial displayed a 9$\times$9 grid of dog images on a white background to emulate a natural environment that may have many competing objects of interest. Gaze-based selection consistently outperforms the other selection modes and is the only one to outperform the users themselves. Significance is indicated at the **$p = 0.01$ and ***$p = 0.001$ levels. Errors bars indicate SE.}
    \label{fig:classification}
\end{figure}

For each trial, a dog image from the test set appeared at one of 25 possible locations in a 5$\times$5 grid, centered relative to the TV and with 8\textdegree{} spacing (Fig.~\ref{fig:classification}). The remaining 80 dog images padded the central image equally on all sides creating a 9$\times$9 grid to emulate natural environments that often have competing objects of interest. Each dog image subtended 8\textdegree{} of visual angle. We added 2\textdegree{} of vertical separation between adjacent images to maintain visually uniform vertical and uniform spacing between dogs.

\paragraph{\textbf{Procedure}}
 We evaluated each selection mode with sets of 25 trials. The order of the modes and trials was randomized, with the central dog image location appearing once in each possible location for each mode. 
 
The procedure for this experiment was largely identical to that of the previous experiment with the only difference being the choice of stimulus. At the start of each selection mode, users were instructed on how to use that mode, were allowed to try practice selections, and then proceeded onto the main trial set. For the gaze-based mode, users calibrated their gaze.

When users clicked away from the waiting screen, we rendered the grid of dog images at one of the possible 25 locations. We highlighted the dog image of interest for the users to select. Once the selection was made, we removed any indication of which dog image was the target and captured the image \Dash users were instructed to remain in place until the image was captured and the wait screen returned. Once all 25 trials for a given selection mode were completed, we proceeded to the next selection mode until all three were completed.

After users evaluated each mode, they identified the dog breeds from the test set based on images alone. We created a quiz that rendered each dog image from the test set alongside a list of the 81 possible dog breeds, the same information that was provided to GPT-4V during analysis (see below). Participants were instructed to identify each dog breed, using a keyboard to navigate through the list and make a selection. We did not enforce a time constraint. They identified each dog breed one at a time until all dogs in the test set were classified.

\paragraph{\textbf{Analysis}}
As in the previous experiment, we recorded the pixel locations of the user's selection (based on gaze or central pixel) in the camera image. We generate a multiscale crop (Sec.~\ref{sec:implementation}) around the selected pixel. This cropped image and a query asking about the dog breed are uploaded to GPT-4V. We then count the number of dog breeds identified by the LMM in each mode and by the user and calculate accuracies.

The query about dog breeds presented to the LMM lists all 81 possible choices of dog breed and asks it to choose one in its response (see Supplement). We included the  breeds in the query for two reasons: first, it sidesteps the issue of what should be done if the response is correct but too generic (e.g., ``terrier'' instead of ``Boston terrier''), and second, the user chooses from the same list. 

We analyzed the classification accuracies with a one-way ANOVA with the independent variable of selection mode/user having four levels. Post hoc tests were conducted as \textit{t}-tests with Bonferroni correction applied to the \textit{p}-values \Dash reported \textit{p}-values have the Bonferroni correction factor applied.

\paragraph{\textbf{Baseline}}
The original set of 25 test dog images was directly uploaded to GPT-4V for baseline analysis with the query, ``Please identify the dog breeds in the following images:''. The dog images were renamed so that the breed name did not appear in the file name. On this set, the LMM classified 84\% of the dog breeds correctly.

To measure the effect of camera quality, we captured multiscale images of the 25 test dog images. The camera was placed 1~m from the study TV to match the users' study distance. We uploaded these images to GPT-4V for classification using the same query as in the study. The LMM classified 64\% of the dog breeds correctly based on these images, showing the degradation in performance due to the camera quality.

We also evaluated the LMM classification baseline when the camera was pointed directly at each dog's face. Faces are salient features and we observed that most of our users fixated on the dogs' faces. With this modification, the LMM classified only 52\% of the dog breeds correctly. While the multiscale images collectively capture the full dog, the part of the dog captured in the most zoomed-in view clearly affects LMM performance.

\paragraph{\textbf{Results}}
The average classification accuracy for each mode can be found in Fig.~\ref{fig:classification}. Without the help of an LMM, users on average identified 22\% of dog breeds. Gaze-based selection has the best accuracy at 40\% and body- and head-based are around 15\%. 

The ANOVA shows a signification effect of mode ($F_{3, 44} = 36.12$, $p < 0.001$). We conduct follow-up \textit{t}-tests for the post hoc analysis, which show that gaze-based selection has significantly higher accuracy than all three other modes (all $p < 0.001$).

From this result we can conclude that gaze-based selection is the only mode that can effectively augment human capabilities. The other selection modes, likely due to their higher selection error, often assume the user is looking at a different object than intended.

\subsection{User Preference}
In this experiment, users experienced the end-to-end contextual AI systems in a real-world environment with the task of qualitatively ranking them along different metrics.

We populated a desk and wall roughly 1~m in front of the user with common objects like plants, tissue boxes, and paintings (see supplement) and users freely inquired about these objects. Using a similar interface to the one described in Sec.~\ref{sec:user_interface}, users selected objects using the gaze-, head-, and body-based modes and spoke a question. The contextual AI would transcribe their speech, upload the query with associated images to GPT-4V, and respond with audio. After the user selected an object, we displayed the multiscale image of that object so the user could adjust behavior for subsequent queries.

The order of selection modes that the user experienced in this setting was randomized per user. Users were asked to perform multiple queries before moving on to the next mode. After experiencing all three modes, users could ask to return to any individual mode. 

We gave users a forced-choice ranking task between the selection modes on different metrics: 1) which was the most natural to select the object of interest, 2) how useful the user could imagine it to be in their daily life, and 3) which provided the fastest response after the selection was made. The third criterion was a control question since the GPT-4V backend was used for all modes. We explained the metrics to users before the start of the evaluation and we reminded them of the criteria when starting each new mode. After experiencing all of the modes, users filled out a questionnaire indicating their rankings along with any additional comments.

\paragraph{\textbf{Results}}
All user preference rankings can be found in Fig.~\ref{fig:prefs} (black dots). There is a clear trend of gaze-based selection being preferred over head- and body-based for naturalness and usefulness. For latency, body-based is perceived as slower, perhaps due to user frustration, but head- and gaze-based are similar.

Friedman tests were significant at the 0.001 level for all preference questions. Wilcoxon signed-rank tests were used for post hoc analysis, with Bonferroni correction applied to the $p$-values. We see that gaze-based is preferred over body-based (all questions $p < 0.01$), head-based is preferred over body-based (naturalness $p < 0.01$, usefulness and latency $p < 0.05$), and gaze-based is preferred over head-based (naturalness $p < 0.01$, usefulness $p < 0.05$).

%% file: sections/5_applications.tex
GazeGPT is most useful in on-the-go environments where users are looking to learn more information about an object in their line of sight. This is particularly useful in scenarios where users don't have time to pull out their phone, launch an LMM app, and take a photo.

For example, when traveling to a foreign country, a user could glance at a street sign, menu, or advertisement and say ``What does this mean?'' and the contextual AI would respond in their ear. When eating at a restaurant, a user could ask which wine could pair best with a meal, acting as a sommelier for the table.

GazeGPT can also be useful for day-to-day life. A user at a grocery store could inquire about the calories or health benefits of a vegetable, and then get recommendations for meals with that vegetable. Out in the forest, a user could inquire whether a mushroom or berry is edible. For those that are colorblind, GazeGPT could be used to answer a question such as ``Do my clothes match?'' Several other real-world scenarios are depicted in Fig.~\ref{fig:gazegpt_applications}.

%% file: sections/6_discussion.tex
Generative AI is already capable of ingesting various types of complex data and outputting reasonable responses \Dash and we expect that to accelerate in coming years. One fundamental challenge that remains, however, is providing these models with context about what is important to the human user. Much like prompt engineering can guide AI models, real-world object selection in the image domain can be an effective method of providing context to user queries, particularly in on-the-go wearable devices.

\begin{figure}[t!]
    \includegraphics[width=\columnwidth]{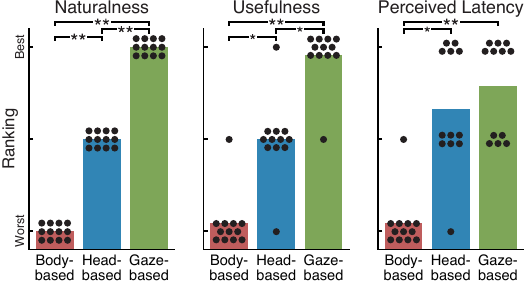}
    \caption{Preference study results. Gaze-based selection is consistently rated as more natural and useful than the others. While the latency of the three modes should be identical, it appears that user frustration with body-based selection affected their perception of latency as well. Significance is indicated at the *$p = 0.05$ and **$p = 0.01$ levels. Each dot represents a user ranking.}
    \vspace{-10pt}
    \label{fig:prefs}
\end{figure}

We evaluated multiple object-selection modes on speed and accuracy. The gaze-based selection mode was \Dash without visual feedback \Dash as accurate as a phone-like system with visual feedback and roughly 3$\times$ faster. We expect even larger differences in speed in real-world use, as a phone in a pocket is even less accessible. Head- and body-based selection modes, while fast, did not maintain the accuracy likely needed for real-world use. 

This was corroborated in a follow-up study in which users used different selection modes to classify dog breeds based on images alone. While  gaze-based selection (40\% correct) outperformed human capabilities (22\% correct) on this task, the head- and body-based selection modes fared worse in comparison (both around 15\% correct). In a user preference evaluation, the gaze-based mode was heavily favored compared to the head- and body-based modes.

\paragraph{\textbf{Limitations and Future Work.}}
We observed that the world-facing camera quality has a significant effect on the classification ability of the system. Occasionally, motion blur or focus error degraded the images. The IMX179-based world facing camera is fairly basic, and well-established techniques such as optical image stabilization and phase detection autofocus could greatly improve performance. At other times, the quality of the LMM's response was degraded simply due to the limited resolution of the IMX179 sensor, which could be mitigated by using a higher resolution sensor.

Secondly, before a GazeGPT system can achieve mass adoption, the LMM response latency will likely need to be reduced significantly. The current GazeGPT system maintains a latency of approximately 5.2~s. This includes speech-to-text (STT), LMM response, and text-to-speech (TTS) models, all processed in the cloud. Bringing these models on device to reduce latency may be feasible for the smaller STT and TTS models, though the LMM will necessarily remain cloud-based in the near future.

Finally, a relatively simple dog-breed-classification task was used to demonstrate how GazeGPT augments human capabilities. Future work could use standard LMM evaluation benchmarks, such as the MMMU (Massive Multi-discipline Multimodal Understanding)~\cite{yue2023mmmu}, to evaluate the benefits of gaze-contingent AI on a broader set of tasks. The MMMU contains 11.5k questions across 30 subjects, however, so an adequate evaluation on many users would be well beyond the scope of this paper.

\paragraph{\textbf{Conclusion}}
Human-centric AI technologies, such as GazeGPT, provide an opportunity for users to directly benefit from the quickly evolving capabilities of generative AI models in everyday tasks. One can only imagine the myriad of creative applications GazeGPT may enable when users start to see the world through the lens of contextual AI.